\documentclass{ws-mpla}
\usepackage[super]{cite}
\usepackage{color}
\usepackage{graphicx}
\usepackage{subfigure}
\usepackage{epstopdf}
\usepackage{hyperref}

\begin{document}

\markboth{Hua-Xing Chen}{Further prediction on the possible double-peak structure of the $X17$ particle}

\catchline{}{}{}{}{}

\title{Further prediction on the possible double-peak structure of the $X17$ particle}

\author{Hua-Xing Chen}
\address{School of Physics, Southeast University, Nanjing 210094, China\\
hxchen@seu.edu.cn}

\maketitle

\pub{Received (Day Month Year)}{Revised (Day Month Year)}

\begin{abstract}
The $X17$ particle, discovered by Krasznahorkay et al. at ATOMKI~\cite{Krasznahorkay:2015iga}, was recently confirmed in the $\gamma \gamma$ invariant mass spectra by Abraamyan et al.  at JINR~\cite{Abraamyan:2023hed}. We notice with surprise and interest that the $X17$ seems to have a double-peak structure. This is in a possible coincidence with our QCD sum rule study of Ref.~\refcite{Chen:2020arr}, where we interpreted the $X17$ as a tetraquark state composed of four bare quarks ($u \bar u d \bar d$), and claimed that ``A unique feature of this tetraquark assignment is that we predict two almost degenerate states with significantly different widths''. These two different tetraquark states are described by two different chiral tetraquark currents $\bar u_L \gamma_\mu d_L~\bar d_L \gamma^\mu u_L$ and $\bar u_L \gamma_\mu d_L~\bar d_R \gamma^\mu u_R$. To verify whether the tetraquark assignment is correct or not, we replace the up and down quarks by the strange quarks, and apply the QCD sum rule method to study the other four chiral tetraquark currents $\bar u_L \gamma_\mu s_L~\bar s_L \gamma^\mu u_L$, $\bar u_L \gamma_\mu s_L~\bar s_R \gamma^\mu u_R$, $\bar d_L \gamma_\mu s_L~\bar s_L \gamma^\mu d_L$, and $\bar d_L \gamma_\mu s_L~\bar s_R \gamma^\mu d_R$. We calculate their correlation functions, and find that non-perturbative QCD effects do not contribute much to them. Our results suggest that there may exist four almost degenerate tetraquark states with masses about $236\sim296$~MeV. Each of these states is composed of four bare quarks, either $u \bar u s \bar s$ or $d \bar d s \bar s$.

\end{abstract}

\ccode{PACS Nos.: 14.40.Rt, 12.38.Lg.}

\section{Introduction}

The hadron is a composite particle made of quarks and gluons bound together by the strong interaction.  Take the proton as an example, it is composed of two valence up quarks and one valence down quark in the traditional quark model, but with the development of QCD as the theory of the strong interaction, we realize that the proton also contains numberless sea quarks and gluons~\cite{ParticleDataGroup:2022pth}. Especially, there can be some sea quarks, whose combination is color singlet. The color confinement, as an essential property of QCD, demands that color-charged particles can not be isolated, and therefore, can not be directly observed in normal conditions below the Hagedorn temperature $T \approx 150$~MeV. Now, a natural question arises: is the color-singlet combination of sea quarks confined in the hadron?

In 2016 Krasznahorkay et al. studied the nuclear reaction $^7$Li$(p,e^+e^-)^8$Be at ATOMKI, and observed an anomaly in the angular correlation of the $e^+e^-$ emission from the excited $^8$Be nucleus~\cite{Krasznahorkay:2015iga}. Later in 2019 and 2022 they observed the same anomaly in the decays of the excited $^4$He and $^{12}$C nuclei~\cite{Krasznahorkay:2017qfd,Krasznahorkay:2019lyl,Krasznahorkay:2022pxs}. This anomaly was interpreted as the signature of a neutral boson with the mass about 17~MeV, called the ``$X17$'' particle, whose observations attracted much interest from theorists. Various explanations were proposed, such as a fifth force~\cite{Feng:2016jff,Feng:2016ysn,Gu:2016ege,Kahn:2016vjr,Fayet:2016nyc,Dror:2017nsg}, dark matter~\cite{Liang:2016ffe,Jia:2016uxs,Ellwanger:2016wfe,Kozaczuk:2016nma,Dror:2017ehi,Hati:2020fzp}, nuclear physics models~\cite{Zhang:2017zap,Veselsky:2020ewb}, QCD axion~\cite{Bauer:2017ris,Alves:2017avw,Kirpichnikov:2020tcf,Dusaev:2020gxi}, and QED meson~\cite{Wong:2020hjc,Wong:2021blz}, etc. We refer to the reviews~\refcite{Barducci:2022lqd,Alves:2023ree} for detailed discussions, and note that the nature of the $X17$ particle is still far beyond our understanding.

In 2020 we investigated the possible assignment of the $X17$ particle as a tetraquark state composed of four bare quarks~\cite{Chen:2020arr}. This state is similar to the color-singlet combination of sea quarks in some aspects. Assuming the two chiral tetraquark currents (detailed explanations will be given later)
\begin{eqnarray}
\nonumber J_{LL} &=& \bar{u}_L \gamma_\mu d_L~\bar{d}_L \gamma^\mu u_L \, ,
\\ \nonumber J_{LR} &=& \bar{u}_L \gamma_\mu d_L~\bar{d}_R \gamma^\mu u_R \, ,
\end{eqnarray}
respectively couple to the two tetraquark states $X_{LL}$ and $X_{LR}$, we used the QCD sum rule method to calculate their masses to be both about $17.3^{+1.4}_{-1.7}$~MeV. We also studied their decay properties in Ref.~\refcite{Chen:2020arr}, but found the width of $X_{LR}$ to be significantly smaller than that of $X_{LL}$. Therefore, we arrived at the unique feature of our tetraquark assignment that ``we predict two almost degenerate states with significantly different widths''.

Very recently, Abraamyan et al. studied the $p$C, $d$C, and $d$Cu collisions at JINR, and observed two enhanced structures in the $\gamma \gamma$ invariant mass spectra at about 17~MeV and 38~MeV~\cite{Abraamyan:2023hed}. This observation confirmed the occurrence of the $X17$ particle at different initial conditions and from different decay channels. We notice with surprise and interest that there seems to exist two peaks in the $\gamma \gamma$ invariant mass spectra at about 17 MeV. This indicates that the $X17$ may have a double-peak structure, which is in a possible coincidence with our QCD sum rule study of Ref.~\refcite{Chen:2020arr}.

Although this may be just a coincidence, it demands us to make further predictions. In this paper we replace the up and down quarks of $J_{LL}$ and $J_{LR}$ by the strange quarks, and construct more chiral tetraquark currents with the quark contents $u \bar u s \bar s$ and $d \bar d s \bar s$. We apply the QCD sum rule method to study these currents. Especially, we find that the non-perturbative QCD effects, such as the quark and gluon condensates, do not contribute much to them. This allows us to choose some abnormal QCD sum rule parameters to arrive at the new prediction that there may exist four almost degenerate tetraquark states with masses about $236\sim296$~MeV. In order to verify whether our tetraquark assignment is correct or not, we propose to search for these states in future particle and nuclear experiments.

This paper is organized as follows. In Sec.~\ref{sec:current} we construct the chiral tetraquark currents with the quark contents $u \bar u s \bar s$ and $d \bar d s \bar s$. We use these currents to perform QCD sum rule analyses in Sec.~\ref{sec:sumrule}, and perform numerical analyses in Sec.~\ref{sec:numerical}. The obtained results are summarized and discussed in Sec.~\ref{sec:summary}.

\section{Chiral tetraquark currents}
\label{sec:current}

A tetraquark current is composed of two quark fields and two antiquark fields. It can be generally written as
\begin{equation}
J(x) = C_{abcd}^{\mu\nu\rho\sigma} ~ \bar{q}^a_\mu(x) ~ \bar{q}^b_\nu(x) ~ q^c_\rho(x) ~ q^d_\sigma(x) \, ,
\label{def:current}
\end{equation}
where $\mu \cdots \sigma$ are Dirac spinor indices, $a \cdots d$ are color indices, $C_{abcd}^{\mu\nu\rho\sigma}$ is the coefficient to provide certain quantum numbers, and the sum over repeated indices is taken.

We have applied the QCD sum rule method to systematically study the scalar and pseudoscalar tetraquark currents in Refs.~\refcite{Chen:2006hy,Chen:2007xr,Dong:2020okt}, and we refer to Ref.~\refcite{Groote:2014pva,Wang:2020cme} for more QCD sum rule studies. For example, the two scalar tetraquark currents ($q=u,d,s$)
\begin{eqnarray}
J_{VV} &=& \bar{q}^a \gamma_\mu q^a~\bar{q}^b \gamma^\mu q^b \, ,
\label{def:JVV}
\\ J_{AA} &=& \bar{q}^a \gamma_\mu \gamma_5 q^a~\bar{q}^b \gamma^\mu \gamma_5 q^b \, ,
\label{def:JAA}
\end{eqnarray}
were combined together in Ref.~\refcite{Chen:2007xr} to explain the light scalar mesons $f_0(500)$, $K_0^*(700)$, $a_0(980)$, and $f_0(980)$ as tetraquark states. Based on these studies, we combined the chiral quark fields
\begin{equation}
\nonumber q_L = {1-\gamma_5\over2} q ~~~~{\rm and}~~~~ q_R = {1+\gamma_5\over2} q \, ,
\end{equation}
to construct four chiral tetraquark currents with the quark content $u \bar u d \bar d$:
\begin{eqnarray}
J^{ud}_{LL} &=& \bar{u}^a_L \gamma_\mu d^a_L~\bar{d}_L^b \gamma^\mu u_L^b \, ,
\label{def:JLLud}
\\ J^{ud}_{LR} &=& \bar{u}^a_L \gamma_\mu d^a_L~\bar{d}_R^b \gamma^\mu u_R^b \, ,
\label{def:JLRud}
\\ J^{ud}_{RL} &=& \bar{u}^a_R \gamma_\mu d^a_R~\bar{d}_L^b \gamma^\mu u_L^b \, ,
\label{def:JRLud}
\\ J^{ud}_{RR} &=& \bar{u}^a_R \gamma_\mu d^a_R~\bar{d}_R^b \gamma^\mu u_R^b \, ,
\label{def:JRRud}
\end{eqnarray}
which were used in Ref.~\refcite{Chen:2020arr} to explain the $X17$ as a possible tetraquark state. Note that the properties of $J^{ud}_{RR}$ and $J^{ud}_{RL}$ are respectively the same as those of $J^{ud}_{LL}$ and $J^{ud}_{LR}$, given that the weak interaction can be neglected here.

We can replace the down quarks by the strange quarks, and construct four chiral tetraquark currents with the quark content $u \bar u s \bar s$:
\begin{eqnarray}
J^{us}_{LL} &=& \bar{u}^a_L \gamma_\mu s^a_L~\bar{s}_L^b \gamma^\mu u_L^b \, ,
\label{def:JLLus}
\\ J^{us}_{LR} &=& \bar{u}^a_L \gamma_\mu s^a_L~\bar{s}_R^b \gamma^\mu u_R^b \, ,
\label{def:JLRus}
\\ J^{us}_{RL} &=& \bar{u}^a_R \gamma_\mu s^a_R~\bar{s}_L^b \gamma^\mu u_L^b \, ,
\label{def:JRLus}
\\ J^{us}_{RR} &=& \bar{u}^a_R \gamma_\mu s^a_R~\bar{s}_R^b \gamma^\mu u_R^b \, .
\label{def:JRRus}
\end{eqnarray}
We can also replace the up quarks by the strange quarks, and construct four chiral tetraquark currents with the quark content $d \bar d s \bar s$:
\begin{eqnarray}
J^{ds}_{LL} &=& \bar{d}^a_L \gamma_\mu s^a_L~\bar{s}_L^b \gamma^\mu d_L^b \, ,
\label{def:JLLds}
\\ J^{ds}_{LR} &=& \bar{d}^a_L \gamma_\mu s^a_L~\bar{s}_R^b \gamma^\mu d_R^b \, ,
\label{def:JLRds}
\\ J^{ds}_{RL} &=& \bar{d}^a_R \gamma_\mu s^a_R~\bar{s}_L^b \gamma^\mu d_L^b \, ,
\label{def:JRLds}
\\ J^{ds}_{RR} &=& \bar{d}^a_R \gamma_\mu s^a_R~\bar{s}_R^b \gamma^\mu d_R^b \, .
\label{def:JRRds}
\end{eqnarray}
Among these currents, the properties of $J^{us}_{RR}/J^{us}_{RL}/J^{ds}_{RR}/J^{ds}_{RL}$ are respectively the same as those of $J^{us}_{LL}/J^{us}_{LR}/J^{ds}_{LL}/J^{ds}_{LR}$.

We shall use the eight currents listed in Eqs.~(\ref{def:JLLus}-\ref{def:JRRds}) to perform QCD sum rule analyses. Before doing this, we note that all the twelve currents listed in Eqs.~(\ref{def:JLLud}-\ref{def:JRRds}) do not have definite $P$-parities, {\it i.e.}, each of them contains both $P=+$ and $P=-$ components. This is much different from the currents investigated in our previous QCD sum rule studies~\cite{Chen:2006hy,Chen:2007xr,Dong:2020okt}, all of which have definite $P$-parities, either $P=+$ or $P=-$. However, the currents need not have definite quantum numbers in principle, {\it e.g.}, the current $J_\mu = \bar d \gamma_\mu u$ coupling to the $\rho^+$ meson does not have a $C$-parity, and the weak neutral current does not have a $P$-parity.

\section{QCD sum rule analyses}
\label{sec:sumrule}

The QCD sum rule method is a powerful non-perturbative method, which has been widely and successfully applied to study hadron properties in the past decades~\cite{Shifman:1978bx,Reinders:1984sr}. In this method we investigate the two-point correlation function
\begin{equation}
\Pi(q^2)\,\equiv\,i\int d^4x~e^{iqx}~\langle 0 | \mathbb{T}\left[ J(x){J^\dagger}(0) \right] | 0 \rangle \, ,
\label{eq:pi}
\end{equation}
at both the hadron and quark-gluon levels. To do this, we take the current $J^{us}_{LL}$ as an example, which couples to the tetraquark state $X^{us}_{LL}$ through
\begin{equation}
\langle 0| J^{us}_{LL} | X^{us}_{LL} \rangle = f_X \, ,
\end{equation}
with $f_X$ the decay constant.

At the hadron level, we express Eq.~(\ref{eq:pi}) in the form of the dispersion relation as
\begin{equation}
\Pi^{us}_{LL}(q^2)=\int^\infty_{s_<}\frac{\rho^{\rm phen}(s)}{s-q^2-i\varepsilon}ds \, ,
\end{equation}
with $s_< = (2m_u + 2m_s)^2$ the physical threshold. We parameterize the phenomenological spectral density $\rho^{\rm phen}(s)$ as one pole dominance for the ground state $X^{us}_{LL}$ together with a continuum contribution:
\begin{eqnarray}
\nonumber \rho^{\rm phen}(s) &=& \sum_n \delta(s - M^2_n) \langle0| J^{us}_{LL} |n\rangle \langle n| {J^{us,\dagger}_{LL}} |0\rangle
\\ &=& f^2_X \delta(s-M^2_X) + \rm{continuum} \, ,
\label{eq:rhophen}
\end{eqnarray}
where $M_X$ is the mass of $X^{us}_{LL}$. After performing the Borel transformation, we obtain
\begin{equation}
\mathcal{B}_{q^2 \rightarrow M_B^2}\Pi^{us}_{LL}(q^2) = f^2_X e^{-M_X^2/M_B^2} + \rm{continuum} \, .
\label{eq:piphen}
\end{equation}

At the quark-gluon level, we insert the current $J^{us}_{LL}$ into Eq.~(\ref{eq:pi}), and calculate it through the method of operator product expansion (OPE):
\begin{eqnarray}
&& \mathcal{B}_{q^2 \rightarrow M_B^2}\Pi^{us}_{LL}(q^2)
\label{eq:piLLus}
\\ \nonumber &=& \int^\infty_{s_<} \Bigg( {s^4 \over 81920 \pi^6} \left( 1 + {4\alpha_s \over \pi} \right)
+ {m_u\langle\bar{q}q\rangle + m_s\langle\bar{s}s\rangle \over 256 \pi^4} s^2
\\ \nonumber && ~~~~~ + {m_u\langle g_s\bar{q}\sigma Gq\rangle + m_s\langle g_s\bar{s}\sigma Gs\rangle \over 256 \pi^4}s \Bigg)~e^{-s/M_B^2}~ds \, ,
\end{eqnarray}
from which we can extract the OPE spectral density $\rho^{\rm OPE}(s)$ to be
\begin{eqnarray}
\nonumber \rho^{\rm OPE}(s) &=& {s^4 \over 81920 \pi^6}  \left( 1 + {4\alpha_s \over \pi} \right)
+ {m_u\langle\bar{q}q\rangle + m_s\langle\bar{s}s\rangle \over 256 \pi^4}s^2
\\ && + {m_u\langle g_s\bar{q}\sigma Gq\rangle + m_s\langle g_s\bar{s}\sigma Gs\rangle \over 256 \pi^4}s \, .
\label{eq:rhoLLus}
\end{eqnarray}
In the above expressions, we have kept the current quark masses up to the $\mathcal{O}(m_q)$ order and performed the OPE calculation up to the twentieth dimension. We have calculated all the leading-order Feynman diagrams, including the perturbative term, the quark condensates, the quark-gluon mixed condensates, the two-/three-gluon condensates, and their combinations. We have partly calculated the next-to-leading-order Feynman diagrams, including the $\mathcal{O}(\alpha_s)$ corrections to the perturbative term, to the quark condensates, to the quark-gluon mixed condensates, and to their combinations.

The results for the three tetraquark currents $J^{us}_{LR/RL/RR}$ are exactly the same:
\begin{equation}
\nonumber \Pi^{us}_{LL}(q^2) = \Pi^{us}_{LR}(q^2) = \Pi^{us}_{RL}(q^2) = \Pi^{us}_{RR}(q^2) \, ,
\end{equation}
and the results for the four tetraquark currents $J^{ds}_{LL/LR/RL/RR}$ can be derived by replacing $m_u \rightarrow m_d$.

Comparing Eq.~(\ref{eq:piphen}) at the hadron level and Eq.~(\ref{eq:piLLus}) at the quark-gluon level, we approximate the continuum contribution by the OPE spectral density $\rho^{\rm OPE}(s)$ above the threshold value $s_0$ to arrive at
\begin{equation}
\nonumber \Pi^{us}_{LL}(s_0, M_B^2) = f^2_X e^{-M_X^2/M_B^2} = \int^{s_0}_{s_<} \rho^{\rm OPE}(s) e^{-s/M_B^2} ds \, ,
\end{equation}
which can be used to calculate the mass $M_X$ through
\begin{equation}
\big[M_X(s_0, M_B^2)\big]^2 = \frac{\int^{s_0}_{s_<} \rho^{\rm OPE}(s)~e^{-s/M_B^2}~s ds}{\int^{s_0}_{s_<} \rho^{\rm OPE}(s)~e^{-s/M_B^2}~ds}\, .
\label{eq:LSR}
\end{equation}

\section{Phenomenological analyses}
\label{sec:numerical}

We use the following values for various QCD parameters at the renormalization scale 1~GeV~\cite{ParticleDataGroup:2022pth,Ovchinnikov:1988gk,Colangelo:1998ga,Jamin:2002ev,Ioffe:2002be,Gimenez:2005nt,Herren:2017osy,Narison:2018dcr}:
\begin{eqnarray}
\nonumber \alpha_s(1~\mbox{GeV}) &=& 0.408 \pm 0.016 \, ,
\\ \nonumber m_u(1~\mbox{GeV}) &=& (2.16^{+0.49}_{-0.26}) \times 1.35 \mbox{ MeV} \, ,
\\ \nonumber m_d(1~\mbox{GeV}) &=& (4.67^{+0.48}_{-0.17}) \times 1.35 \mbox{ MeV} \, ,
\\ \nonumber m_s(1~\mbox{GeV}) &=& (93^{+11}_{-~5}) \times 1.35 \mbox{ MeV} \, ,
\\ \langle \alpha_s GG\rangle &=& (6.35 \pm 0.35) \times 10^{-2} \mbox{ GeV}^4 \, ,
\label{eq:condensates}
\\ \nonumber \langle\bar qq \rangle &=& -(0.24 \pm 0.01)^3 \mbox{ GeV}^3 \, ,
\\ \nonumber \langle\bar ss \rangle &=& (0.8\pm 0.1)\times \langle\bar qq \rangle \, ,
\\ \nonumber \langle g_s \bar q \sigma G q\rangle &=& - (0.8 \pm 0.2) \times \langle\bar qq\rangle \, ,
\\ \nonumber \langle g_s \bar s\sigma G s\rangle &=& - (0.8 \pm 0.2) \times \langle\bar ss\rangle \, .
\end{eqnarray}
Besides, there are two QCD sum rule parameters: the threshold value $s_0$ and the Borel mass $M_B$. In Ref.~\refcite{Chen:2007xr} we took $s_0 = 1.7$~GeV$^2$ and $M_B = 0.8$~GeV when interpreting the light scalar mesons $a_0(980)$ and $f_0(980)$ as two tetraquark states with the quark content $q\bar q s \bar s$ ($q=u,d$). Still using these two values, we numerically calculate Eq.~(\ref{eq:piLLus}) to be:
\begin{eqnarray}
&& \Pi^{us}_{LL}(s_0 = 1.7~{\rm GeV}^2, \, M_B = 0.8~{\rm GeV})
\label{eq:piLLusnum}
\\ \nonumber &=& \Big( 4.3~[{\rm LO}] + 2.2~[{\rm NLO}] - 14.9~[{\rm D}^4] + 14.0~[{\rm D}^6] \Big)
\\ \nonumber && ~~~~~~~~~~~~~~~~~~~~~~~~~~~~~~~~~~~~~~~~~~~~~~ \cdot 10^{-9}~{\rm GeV}^{10} \, .
\end{eqnarray}
In the above expression, the $\mathcal{O}(\alpha_s)$ correction to the perturbative term ([NLO]) is half of the leading-order perturbative term ([LO]); the two non-perturbative terms of $[{\rm D}={4}]$ and $[{\rm D}={6}]$ are both three times larger than the leading-order perturbative term, but their summation is much smaller; the higher-order non-perturbative terms of $[{\rm D} > 6]$ all vanish, which makes the OPE convergence quite good.

%
\begin{figure*}[hbtp]
\begin{center}
\includegraphics[width=0.4\textwidth]{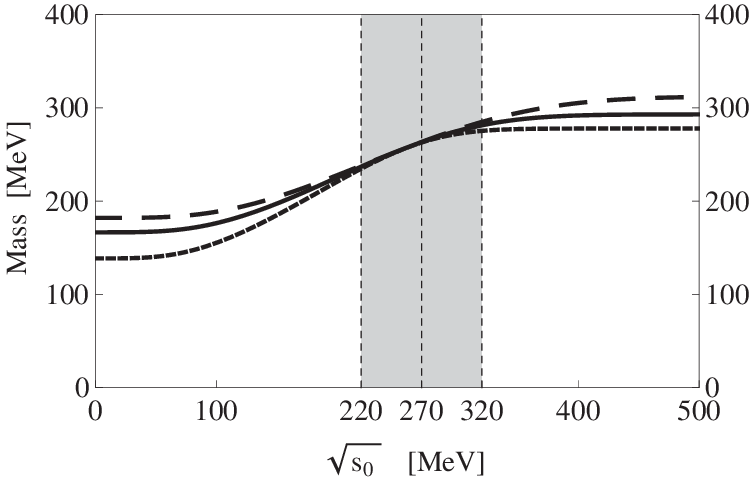}
~~~~~~~~~~
\includegraphics[width=0.4\textwidth]{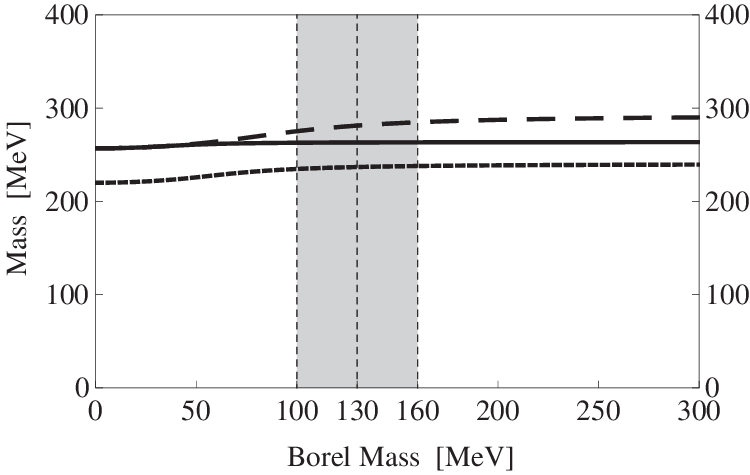}
\caption{The mass of the tetraquark state $X^{us}_{LL}$, extracted from the chiral tetraquark current $J^{us}_{LL}$, with respect to the threshold value $s_0$ (left) and the Borel mass $M_B$ (right). In the left panel the short-dashed/solid/long-dashed curves are obtained by setting $M_B = 100/130/160$~MeV, respectively. In the right panel the short-dashed/solid/long-dashed curves are obtained by setting $\sqrt{s_0} = 220/270/320$~MeV, respectively.}
\label{fig:JLLmass}
\end{center}
\end{figure*}
%

For comparisons, we also use the ``normal'' tetraquark currents $J_{VV}$ and $J_{AA}$, as defined in Eqs.~(\ref{def:JVV}) and (\ref{def:JAA}), to perform QCD sum rule analyses. Setting their quark content to be $u\bar u s \bar s$, {\it i.e.},
\begin{eqnarray}
J^{us}_{VV} &=& \bar{u}^a \gamma_\mu s^a~\bar{s}^b \gamma^\mu u^b \, ,
\\ J^{us}_{AA} &=& \bar{u}^a \gamma_\mu \gamma_5 s^a~\bar{s}^b \gamma^\mu \gamma_5 u^b \, ,
\end{eqnarray}
we numerically calculate their correlation functions to be:
\begin{eqnarray}
&& {1\over4}\cdot\Pi^{us}_{VV}(s_0 = 1.7~{\rm GeV}^2, \, M_B = 0.8~{\rm GeV})
\label{eq:piVVusnum}
\\ \nonumber &=& \Big( 4.3~[{\rm LO}] + 3.5~[{\rm D}^4] + 135.4~[{\rm D}^6] - 247.3~[{\rm D}^8]
\\ \nonumber && ~~~~~~~~~~~~ + 48.1~[{\rm D}^{10}] + 28.4~[{\rm D}^{12}] \Big)
\cdot 10^{-9}~{\rm GeV}^{10} ,
\\
&& {1\over4}\cdot\Pi^{us}_{AA}(s_0 = 1.7~{\rm GeV}^2, \, M_B = 0.8~{\rm GeV})
\label{eq:piAAusnum}
\\ \nonumber &=& \Big( 4.3~[{\rm LO}] - 33.3~[{\rm D}^4] - 107.5~[{\rm D}^6] + 247.3~[{\rm D}^8]
\\ \nonumber && ~~~~~~~~~~~~ + 42.0~[{\rm D}^{10}] + 83.8~[{\rm D}^{12}] \Big)
\cdot 10^{-9}~{\rm GeV}^{10} .
\end{eqnarray}
In the above expressions, the leading-order perturbative term ([LO]) has been rescaled to be the same as Eq.~(\ref{eq:piLLusnum}); the non-perturbative terms of $[{\rm D}={6 \sim 12}]$ are all much larger than the perturbative term; we have kept the current quark masses up to the $\mathcal{O}(m_q)$ order and performed the OPE calculation up to the twelfth dimension, while the calculations of the $[{\rm D} > 12]$ terms (some are non-zero) are not so easy that these terms are not investigated in the present study.

Comparing Eq.~(\ref{eq:piLLusnum}) and Eqs.~(\ref{eq:piVVusnum}-\ref{eq:piAAusnum}), we find that the contributions of the non-perturbative terms to the ``normal'' tetraquark currents $J^{us}_{VV}$ and $J^{us}_{AA}$ are much larger than their contributions to the chiral tetraquark current $J^{us}_{LL}$. This allows us to choose an abnormally small $s_0$ to perform phenomenological analyses. To extract the mass of $X^{us}_{LL}$ from $J^{us}_{LL}$ through Eq.~(\ref{eq:LSR}), we need to find proper working regions for the threshold value $s_0$ and the Borel mass $M_B$. The first criterion is to investigate the convergence of Eq.~(\ref{eq:piLLus}). Since the higher-order terms of $[{\rm D} > 6]$ all vanish, its convergence is quite good. The second criterion is to investigate the one-pole-dominance assumption by requiring the pole contribution to be larger than 30\%:
\begin{equation}
\mbox{PC} \equiv \left|\frac{ \Pi^{us}_{LL}(s_0, M_B^2) }{ \Pi^{us}_{LL}(\infty, M_B^2) }\right| \geq 30\% \, .
\label{eq:pole}
\end{equation}
Besides, there exists a natural criterion that $\sqrt{s_0}$ should be larger than the mass of $X^{us}_{LL}$. Accordingly, the smallest threshold value we can take is about $\sqrt{s_0} = 270$~MeV. We use Eq.~(\ref{eq:pole}) to further determine $M_B \leq 130$~MeV. Using these two values, we calculate the mass of $X^{us}_{LL}$ to be
\begin{equation}
M^{us}_{LL} = 263^{+23}_{-27}~{\rm MeV} \, ,
\label{eq:massus}
\end{equation}
where the uncertainty is estimated by setting $\sqrt{s_0} = 270 \pm 50$~MeV and $M_B = 130\pm30$~MeV as well as taking into account the uncertainties of various QCD parameters listed in Eqs.~(\ref{eq:condensates}). Note that the two central values, $\sqrt{s_0} = 270$~MeV is only slightly larger than $M^{us}_{LL} = 263$~MeV, which is valid only when the width of this state is quite small. We show the mass $M^{us}_{LL}$ in Fig.~\ref{fig:JLLmass} with respect to $s_0$ and $M_B$. Its dependence on $M_B$ is rather weak, but it does depend on $s_0$ to some extent. This suggests that the above mass calculation should be treated with caution, given that the QCD sum rule method is actually a non-perturbative method.

The masses extracted from the three tetraquark currents $J^{us}_{LR/RL/RR}$ are the same as Eq.~(\ref{eq:massus}):
\begin{equation}
\nonumber M^{us}_{LL/LR/RL/RR} = 263^{+23}_{-27}~{\rm MeV} \, ,
\end{equation}
and the masses extracted for the four tetraquark currents $J^{ds}_{LL/LR/RL/RR}$ are similar:
\begin{equation}
\nonumber M^{ds}_{LL/LR/RL/RR} = 272^{+24}_{-27}~{\rm MeV} \, .
\end{equation}
Considering that the properties of $J^{us}_{RR}/J^{us}_{RL}/J^{ds}_{RR}/J^{ds}_{RL}$ are respectively the same as those of $J^{us}_{LL}/J^{us}_{LR}/J^{ds}_{LL}/J^{ds}_{LR}$, our results suggest that there may exist two almost degenerate tetraquark states with the quark content $u\bar u s \bar s$ and two almost degenerate tetraquark states with the quark content $d\bar d s \bar s$, so altogether there may exist four almost degenerate tetraquark states with masses about $236\sim296$~MeV.

\section{Summary and Discussions}
\label{sec:summary}

Very recently, Abraamyan et al. observed an enhanced structure in the $\gamma \gamma$ invariant mass spectra at about 17~MeV~\cite{Abraamyan:2023hed}, which confirmed the existence of the $X17$ particle at different initial conditions and from different decay channels. As shown in Fig.~4, Fig.~5, and Fig.~7 of Ref.~\refcite{Abraamyan:2023hed}, the $X17$ particle seems to have a double-peak structure. Correspondingly, we have applied the QCD sum rule method to interpret the $X17$ as a tetraquark state and claimed that ``A unique feature of this tetraquark assignment is that we predict two almost degenerate states with significantly different widths''~\cite{Chen:2020arr}. This may be just a coincidence, but demands us to do more theoretical studies and make further predictions.

The two different tetraquark states proposed in Ref.~\refcite{Chen:2020arr} are described by two different chiral tetraquark currents $J^{ud}_{LL}$ and $J^{ud}_{LR}$, both of which have the quark content $u \bar u d \bar d$. In this paper we replace the up and down quarks by the strange quarks, and apply the QCD sum rule method to study more chiral tetraquark currents with the quark contents $u \bar u s \bar s$ and $d \bar d s \bar s$. The obtained results suggest that there may exist four almost degenerate tetraquark states with masses about $236\sim296$~MeV, each of which is composed of four bare quarks. To end this paper, we note again that the nature of the $X17$ particle is far beyond our understanding. Since it was observed in the nuclear experiments performed at ATOMKI~\cite{Krasznahorkay:2015iga,Krasznahorkay:2017qfd,Krasznahorkay:2019lyl,Krasznahorkay:2022pxs} and JINR~\cite{Abraamyan:2023hed}, its relevant experimental and theoretical studies can probably improve our understanding on the internal structure of hadrons, on the non-perturbative properties of QCD, and on the color confinement.

\section*{Acknowledgments}

This project is supported by the National Natural Science Foundation of China under Grant No.~12075019,
the Jiangsu Provincial Double-Innovation Program under Grant No.~JSSCRC2021488,
and
the Fundamental Research Funds for the Central Universities.

\bibliographystyle{elsarticle-num}
\bibliography{ref}

\end{document}